\documentclass[aps,reprint,groupedaddress]{revtex4-2}

\usepackage{amsmath}
\usepackage{mathrsfs}
\usepackage{txfonts}
\usepackage{graphicx}

\usepackage[english]{babel}

\begin{document}

\title{Frustrations on Decorated Square Lattice in Ising Model}

\author{F. A. Kassan-Ogly}
\email{Felix.Kassan-Ogly@imp.uran.ru}
%http://orcid.org/0000-0002-0419-0433

\author{A. V. Zarubin}
\email{Alexander.Zarubin@imp.uran.ru}
%http://orcid.org/0000-0001-9193-7701

\affiliation{M. N. Mikheev Institute of Metal Physics of Ural Branch of Russian Academy of Sciences, S. Kovalevskoy Street 18, 620108 Ekaterinburg,
Russia}

\begin{abstract}
In this paper, we performed the comprehensive studies of frustration properties in the Ising model on a decorated square lattice in the framework of an exact analytical approach based on the Kramers--Wannier transfer matrix method. The assigned challenge consisted in finding rigorous formulas for residual entropies of all possible frustrations without exception and elucidating the influence of frustrations on the behavior of the thermodynamic functions of the system under discussion. The accomplished results turned out to be very rich and intricate, especially in comparison with the original undecorated square lattice, in which frustration properties are completely absent. Such an abundance of properties is due to the multiparametric problem of decorated lattice with four exchange interactions $J_{x}$, $J_{y}$, $J_{dx}$ and $J_{dy}$ taking on arbitrary continuous values and different signs: positive (ferromagnetic) and negative (antiferromagnetic), as well as with two discrete variables $d_{x}$ and $d_{y}$ (multiplicities of decorating spins in both lattice directions), corresponding to infinite natural series of numbers.
\end{abstract}
\maketitle

\section{Introduction}

Frustrations in the solid state physics are usually formulated as the impossibility of simultaneous minimization of all the Hamiltonian terms in the presence of competing interactions, which leads to the degeneracy of the system ground state with nonzero entropy at zero temperature.

The conception ``frustrations'' was brought in magnetism by Gerard Toulouse in 1977 \citep{Toulouse:1977:1,Toulouse:1977:2}. It is worthy to notice that actually the phenomenon of \emph{magnetic} frustrations was first discovered by Gregory H. Wannier in 1950 \citep{Wannier:1950,Wannier:1973} in the Ising model on an antiferromagnetic triangular lattice and later on by Kenzi Kan\^{o} and Shigeo Naya on an antiferromagnetic kagome lattice in 1953 \citep{Kano:1953}. In both papers there had been established the absence of a phase transition and the exact nonzero values of entropy in the ground state of a system (at $T=0$) were obtained for the first time in the history of magnetism 
\[
S_{\text{triangular}}^{\circ}=\frac{2}{\pi}\intop_{0}^{\pi/3}\ln(2\cos\omega)\,d\omega\approx0.323\,07,
\]
\begin{multline*}
S_{\text{kagome}}^{\circ}=\frac{1}{24\pi^{2}}\intop_{0}^{2\pi}\intop_{0}^{2\pi}\ln[21-4(\cos\omega_{1}+\cos\omega_{2}\\
+\cos(\omega_{1}+\cos\omega_{2}))]\,d\omega_{1}\,d\omega_{2}\approx0.501\,83.
\end{multline*}

However, the triangular Ising antiferromagnet was not the first condensed-matter system identified as having nonzero value of entropy at $T=0$. In fact, the first such system was not even a magnetic one. Earlier in 1933, William Giauque (Chemistry Nobel Prize, 1949) et al. determined that common water ice possesses such entropy~\citep{Giauque:1933}. This result was explained by Linus Pauling (Chemistry Nobel Prize, 1954) in 1935 \citep{Pauling:1935} in terms of a macroscopic number of proton configurations in water ice arising from the mismatch between the crystalline symmetry of ice and the local hydrogen bonding requirement of the water molecule. Probably in this paper, the nonzero value of entropy at $T=0$ was for the first time called \emph{residual entropy}, from whence it is usable all over the frustration literature and because of the analogy with water ice, many magnetic systems with frustrations have been coined the name \emph{spin ice}. 

One of the most spectacular achievements in the theory of entropy is the problem of water ice on a 2D square lattice that was solved by Elliott H.~Lieb \citep{Lieb:1967} in 1967 by the extremely efficient Kramers--Wannier transfer-matrix method that had been developed in Refs.~\citep{Kramers:1941:1,Kramers:1941:2}. Lieb has obtained the exact expression for the residual entropy $(4/3)^{3/2}$ known as \emph{Lieb's square ice constant} and that is now enters in the list of top mathematical constants, ever discovered by mankind. This constant is represented in the form of the infinite fraction
\[
\left(\frac{4}{3}\right)^{3/2}=1+\frac{1}{1+\dfrac{1}{1+\dfrac{1}{5+\dfrac{1}{1+\dfrac{1}{4+...}}}}}.
\]

Much earlier Lars Onsager \citep{Onsager:1944} used the Kramers--Wannier transfer-matrix method, where for the first time an exact solution for the Helmholtz free energy on a square lattice in the Ising model was obtained, which actually defined the boundary of demarcation between the previous naive and very simplified theories of the mean (molecular) field type and the modern era of the study of critical phenomena in statistical physics with accompanying peculiar features of the behavior of thermodynamic quantities.

In recent years, under the influence of an increasing stream of experimental research, a rapidly developing field has emerged in the theory of magnetism, associated with frustrations in magnets of various dimensions in multifarious lattices and diversified models. It can be instantiated so much as the major conference in the frustrations field, ``Highly Frustrated Magnetism'', that appeared in 2000, now takes place biannually, show very rapid growth in participant number and the number of represented reports covering most topical aspects of magnetism such as: magnetic order in geometrically frustrated magnets; spin glasses and random magnets; quantum frustrated magnetism and spin liquids; itinerant frustrated systems; exotic phenomena induced by macroscopic degeneracy; artificial and molecular frustrated magnets.

Notwithstanding the frustrated magnetism has become an extremely developing field of research and one may anticipate further increases over the coming years, it should be stressed that the concept of frustration is often used, misused and even abused in the literature, and the widespread utilizations of this concept in various disciplines lead to many contradictions and misconceptions especially regarding the residual entropy.

The narration of a state-of-the-art of the vast and extremely quickly developing field ``Frustrated Magnetism'' can be found in two excellent books: ``Introduction to Frustrated Magnetism'', eds. C.~Lacroix, P.~Mendels, and F.~Mila \citep{Lacroix:2011} and ``Frustrated spin systems'', ed. H.~T.~Diep~\citep{Diep:2020}.

The concept of a decorated lattice, related to the magnetic Ising model, was originally proposed in 1951 by Itiro Sy\^{o}zi in attempting to exactly solve the Ising model on the kagome lattice~\citep{Syozi:1951}. It consists in introducing an extra (decorating) spin at the middle point on every bond between the sites of the original lattice (nodal spins). Actually, Sy\^{o}zi introduced a new transformation (the so-called decoration-iteration transformation) by means of which the thermodynamic properties of the decorated lattice can be deduced from those of the original undecorated one.

Although the rigorous solution to kagome lattice was not obtained in~Ref.~\citep{Syozi:1951} nevertheless this concept gave powerful impetus to further generalizations and applications of this concept to various lattices and other models. It should be noted here that the kagome lattice problem was solved two years later by Kenzi Kan\^{o} and Shigeo Naya without resorting to the decoration concept~\citep{Kano:1953}.

A great deal of research interest aimed at exactly solvable decorated models has resulted in a rather impressive list of existing literature that continues to replenish every year. This is determined on the one hand by the fact that the vast majority of real structures are decorated, and on the other hand by a plethora of new phenomena and effects that are only intrinsic to the decorated systems and lacking in the undecorated ones. 

The decoration-iteration transformation was applied to a special case of the so-called semi-decorated square lattice~\citep{Syozi:1968,Yeh:1973}, which means that the bonds of only one direction were decorated by an extra spin leaving another direction undecorated. 

The decoration-iteration transformation was generalized to multiple decorations with arbitrary number of extra spins on every bond of the original lattice~\citep{Miyazima:1968,Syozi:1972,Nakano:1969}.

The Sy\^{o}zi's simple decoration-iteration transformation had undergone various alterations. Itiro Sy\^{o}zi and Huzio Nakano \citep{SyoziNakano:1968} introduced even more complicated modification of decoration, namely, the decoration of decoration interaction, called by them ``super-decoration''. Keiji Yamada \citep{Yamada:1969} considered some planar lattices decorated with higher Ising spins, which can take values greater than two. Yasuhiro Kasai, Sasuke Miyazima and Itiro Sy\^{o}zi \citep{Kasai:1969} suggested one more peculiar version of decoration~-- the so-called double-bond decoration.

One more important achievement in the theory of entropy should be specially noted. Despite the fact that the exact solution of the Ising model in the presence of an external magnetic field has not yet been obtained on any of the planar lattices (including the square one), nevertheless, in the paper by B.~D.~Sto\v{s}i\'{c}, T.~Sto\v{s}i\'{c}, I.~P.~Fittipaldi, and J.~J.~P.~Veerman \citep{Stosic:1997}, an accurate analytical expression was obtained for the residual entropy in the Ising model on the square lattice with allowance of only nearest-neighbor antiferromagnetic exchange interaction $J<0$ in the frustration field $H=4J$. This residual entropy
\[
S^{\circ}=\ln\left[\frac{2(1+\sqrt{2})}{1+\sqrt{5}}\right]
\]
is represented as the natural logarithm of a ratio of the so-called metal ratios known in the wonderful world of number mathematics, namely, the silver ratio and the golden ratio
\[
1+\sqrt{2},\quad\frac{1+\sqrt{5}}{2}.
\]
Inter alia, the concept of golden ratio appeared more than two and a half millenniums ago.

\begin{figure}[htb]
\centering \includegraphics{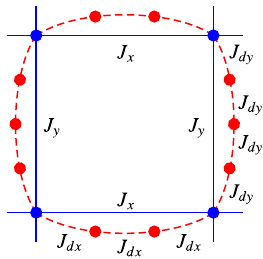}
\caption{Square lattice doubly decorated in $x$-direction and triply decorated in $y$-direction. Blue circles~-- nodal spins; red circles~-- decorating spins. Blue solid lines~-- direct exchange interaction; red dashed lines~-- decorating exchange interaction}
\label{fig:01}
\end{figure}

In this paper we confine ourselves to investigating the simply decorated square Ising lattice (see Fig.~\ref{fig:01}) with arbitrary multiplicities of decorating spins in both lattice directions, focusing on finding all possible cases of frustrations and deriving the exact expressions for corresponding residual entropies.

\section{Thermodynamic functions of the models}

The Hamiltonian for the decorated square lattice has the form
\begin{multline*}
\mathscr{H}=-\sum_{x=1}^{N_{x}}\sum_{y=1}^{N_{y}}\Biggl(J_{x}\sigma_{x,y}\sigma_{x+1,y}+J_{y}\sigma_{x,y}\sigma_{x,y+1}\\
+\sum_{i=1}^{d_{x}+1}J_{dx}\sigma_{x,y}\sigma_{x+i,y}+\sum_{j=1}^{d_{y}+1}J_{dy}\sigma_{x,y}\sigma_{x,y+j}\Biggr),
\end{multline*}
here $J_{x}$ is the exchange interaction between atomic spins at the nodal sites of the nearest neighbors of the original undecorated lattice in the $x$-direction (direct interaction) and, accordingly, $J_{y}$ is the direct interaction in the $y$-direction.
$J_{dx}$ is the exchange interaction both between neighboring decorating spins and between neighboring decorating and nodal spins in the $x$-direction and $J_{dy}$ is accordingly the similar interaction in the $y$-direction.
All the symbols $\sigma$ acquire the values $\pm 1$. $N_{x}$ and $N_{y}$ are the numbers of original undecorated lattice nodes in the $x$- and $y$-directions, and $d_{x}$ and $d_{y}$ are the numbers of decorating spins in the unit cell of the lattice in the $x$- and $y$-directions (decoration multiplicities). 

Using the conception of the decoration-iterative transformation proposed in Ref.~\citep{Syozi:1951}, we obtained analytical expression for the principal (the only one maximal real) eigenvalue of the Kramers--Wannier transfer-matrix for arbitrary decorated square lattice in the most convenient canonical form
\begin{multline*}
\ln\frac{\lambda}{2}=\frac{1}{8\pi^{2}(1+d_{x}+d_{y})}\intop_{0}^{2\pi}\intop_{0}^{2\pi}\ln[C_{x}C_{y}\\
-S_{x}D_{y}\cos\alpha-D_{x}S_{y}\cos\beta]\,d\phi_{x}\,d\phi_{y},
\label{eq:SQD:L1}
\end{multline*}
where
\begin{multline*}
C_{i}=\frac{1}{2}e^{2\frac{J_{i}}{T}}\left(\cosh^{d_{i}+1}\frac{J_{i}^{\prime}}{T}+\sinh^{d_{i}+1}\frac{J_{i}^{\prime}}{T}\right)^{2}\\
+\frac{1}{2}e^{-2\frac{J_{i}}{T}}\left(\cosh^{d_{i}+1}\frac{J_{i}^{\prime}}{T}-\sinh^{d_{i}+1}\frac{J_{i}^{\prime}}{T}\right)^{2},
\end{multline*}
\begin{multline*}
S_{i}=\frac{1}{2}e^{2\frac{J_{i}}{T}}\left(\cosh^{d_{i}+1}\frac{J_{i}^{\prime}}{T}+\sinh^{d_{i}+1}\frac{J_{i}^{\prime}}{T}\right)^{2}\\
-\frac{1}{2}e^{-2\frac{J_{i}}{T}}\left(\cosh^{d_{i}+1}\frac{J_{i}^{\prime}}{T}-\sinh^{d_{i}+1}\frac{J_{i}^{\prime}}{T}\right)^{2},
\end{multline*}
\[
D_{i}=\left(\cosh^{d_{i}+1}\frac{J_{i}^{\prime}}{T}\right)^{2}-\left(\sinh^{d_{i}+1}\frac{J_{i}^{\prime}}{T}\right)^{2},\quad i=x,\ y,
\]
and where $T$ is the absolute temperature.

Given the principal eigenvalue ($\lambda $) of the Kramers--Wannier transfer-matrix, we can calculate the specific entropy
\[
S=\ln\lambda+\frac{T}{\lambda}\frac{\partial\lambda}{\partial T},
\]
as well as the magnetic specific heat capacity
\[
C=2\frac{T}{\lambda}\frac{\partial\lambda}{\partial T}+\frac{T^{2}}{\lambda}\frac{\partial^{2}\lambda}{\partial T^{2}}-\frac{T^{2}}{\lambda^{2}}\left(\frac{\partial\lambda}{\partial T}\right)^{2}.
\]

We also generalized the spontaneous magnetization determined as the square root from pairwise spin-spin correlation function at distance between spins going to infinity (see, for example, Refs.~\citep{Montroll:1963,Baxter:2011}). The spontaneous magnetization of the decorated square lattice that is generalized from the famous Onsager’s expression for undecorated lattice (formula) has the form
\[
M=\left(1-\frac{D_{x}^{2}D_{y}^{2}}{S_{x}^{2}S_{y}^{2}}\right)^{1/8}.
\]
The principal eigenvalue of the Kramers--Wannier transfer-matrix and the spontaneous magnetization of the square lattice with arbitrary decoration multiplicities were obtained in our previous papers, for example~\citep{Kassan-Ogly:2023,Zarubin:2020}.

We have revealed that in the model, proposed in our paper with all possible parameters and above adopted restrictions, only two mechanisms can give rise to frustrations.

First mechanism (free-spins). The decorating interaction $J_{dx}$ ($J_{dy}$) is equal to zero. This is the case of the so-called free (unbound) spins. In this mechanism none restrictions are imposed on the direct interaction $J_{x}$ ($J_{y}$.

Second mechanism (competing interactions). The direct exchange interactions between nodal spins occupying the sites of original simple square lattice $J_{x}$ ($J_{y}$) and the exchange interactions between decorating spins and between decorating and nodal spins $J_{dx}$ ($J_{dy}$) that are in a competition with the correspondent direct interactions. The interactions to be able to generate frustrations should subject to the following conditions, which depend on the interactions signs and the parities of decoration multiplicities $d_{x}$ ($d_{y}$). At odd multiplicity $d_{x}$ the interaction $J_{x}$ must be only negative (antiferromagnetic) and the corresponding interaction $J_{dx}$ may be both negative (antiferromagnetic) and positive (ferromagnetic). At even multiplicity $d_{x}$ the interaction $J_{x}$ may be negative and the corresponding interaction $J_{dx}$ must be positive, or contrariwise, $J_{x}$ positive and $J_{dx}$ negative. In any case, the values of competing interactions $J_{x}$ and $J_{dx}$ should comply with the following requirement
\begin{equation}
|J_{x}|\geqslant|J_{dx}|.
\label{eq:reqx}
\end{equation}

These two mechanisms can operate either self-reliantly or jointly, generating seven possible regimes of frustrations.

\section{First regime. Free spins in semi-decorated lattice}
\label{sec:1r}

Let us consider the simplest regime when the first decorating mechanism is in operation along one direction of the lattice ($x$-direction without loss of generality) and there are no decorations along the other $y$-direction (i.e. $d_{y}=0$, $J_{dy}=0$). This regime is convenient to be called \emph{semi-decorated square lattice with free spins}. In this regime we have multiparameter problem with two exchange interactions $J_{x}$ and $J_{y}$ taking on arbitrary continuous values and arbitrary signs: either positive (ferromagnetic) or negative (antiferromagnetic), as well as with one discrete variable $d_{x}$ (multiplicity of decorating spins in the $x$-direction), corresponding to an infinite natural series of numbers.

\begin{figure}[htb]
\centering \includegraphics{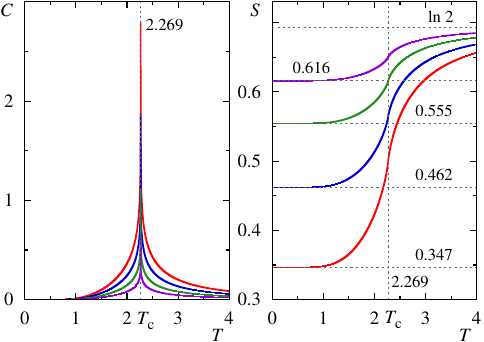}
\caption{Heat capacity (left panel) and entropy (right panel) in semi-decorated square lattice with free spins at $d_{y}=0$, $J_{dy}=0$, $J_{y}=+1$, $J_{x}=+1$, and $J_{dx}=0$. Red curves~-- $d_{x}=1$, blue curves~-- $d_{x}=2$, green curves~-- $d_{x}=4$, violet curves~-- $d_{x}=8$}
\label{fig:02}
\end{figure}

\begin{figure}[htb]
\centering \includegraphics{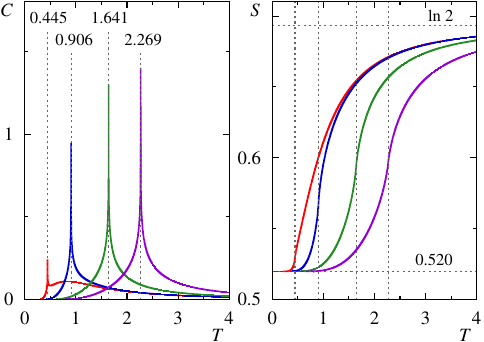}
\caption{Heat capacity (left panel) and entropy (right panel) in semi-decorated square lattice with free spins at $d_{y}=0$, $J_{dy}=0$, $J_{y}=+1$, $J_{dx}=0$ and $d_{x}=3$. Red curves~-- $J_{x}=+0.005$, blue curves~-- $J_{x}=+0.1$, green curves~-- $J_{x}=+0.5$, violet curves~-- $J_{x}=+1$}
\label{fig:03}
\end{figure}

For the comprehensive solution of the whole problem, we performed the overall variation of all above-mentioned parameters. As fairly representative examples, Figures~\ref{fig:02} and \ref{fig:03} present some results of calculating the temperature dependences of heat capacity and entropy on a semi-decorated square lattice with free spins in one direction at varying the decoration multiplicities, the values, as well as signs of exchange interactions. Namely, in Figure~\ref{fig:02} only the multiplicity of decoration $d_{x}$ changes at fixed exchange interactions, and, as a consequence, all the phase-transition points coincide while the residual entropy acquire different values.

In Figure~\ref{fig:03}, on the contrary, only the exchange interaction $J_{x}$ changes at fixed decoration multiplicity $d_{x}$ and, as a consequence, the phase-transition points acquire different values while the residual entropy retains invariable.

From numerous computations, a part of which are depicted in Figs.~\ref{fig:02} and \ref{fig:03}, we arrive at the following conclusions. The first mechanism, in which the decorating interaction is equal to zero, operates equally without any restrictions on the signs and values of direct interactions $J_{x}$ and $J_{y}$. The residual entropy in all possible cases only depends on multiplicity $d_{x}$ by the universal formula
\begin{equation}
S_{\text{free\ semi}}^{\circ}=\frac{d_{x}\ln2}{1+d_{x}},
\label{eq:13}
\end{equation}
irrespective of the other parameters, including even the case of \emph{both ferromagnetic} interactions.

One remark is worthy to specially be noticed that in all the cases the frustrations, confirmed by nonzero residual entropy, are always accompanied by the phase transition evidenced by a sharp Onsager's peak in the heat capacity (sometimes splitting with additional dome-shaped maximum), which demythologizes (refutes) the widespread opinion (myth) about frustrations and phase transitions as mutually exclusive phenomena. Since in the free-spins regime the phase-transition points are determined by Onsager's formulas it allows only one phase transition disabling multiple transitions at all. The second worthy of note remark involves the other myth about impossibility of frustrations in the systems with ferromagnetic exchange interactions.

\section{Second regime. Competing interactions in semi-decorated lattice}
\label{sec:2r}

Let us consider now the next regime in which the second decorating mechanism operates along one direction of the lattice ($x$-direction without loss of generality) and again there are no decorations along the other $y$-direction. This regime should be called \emph{semi-decorated square lattice with competing direct interaction and interaction between decorating spins}. In this regime one more variable parameter is subjoined to the multiparametric problem, namely, the decorating exchange interaction $J_{dx}$.

The interactions to be competing should satisfy the following conditions that depend on the interactions signs and the parity of the decoration multiplicity $d_{x}$.

At even decoration multiplicity $d_{x}$ the interaction $J_{dx}$ may be negative (antiferromagnetic) and the corresponding interaction $J_{x}$ must be positive (ferromagnetic), or contrariwise: $J_{dx}$ positive and $J_{x}$ negative.

At odd decoration multiplicity $d_{x}$ the interaction $J_{dx}$ may be both positive (ferromagnetic) and negative (antiferromagnetic). However, in both cases, the direct exchange interaction should only be negative (antiferromagnetic).

For frustrations to appear, the values of interactions $J_{x}$ and $J_{dx}$ should submit to the following interrelation (see Eq.~(\ref{eq:reqx})).

\begin{figure}[htb]
\centering \includegraphics{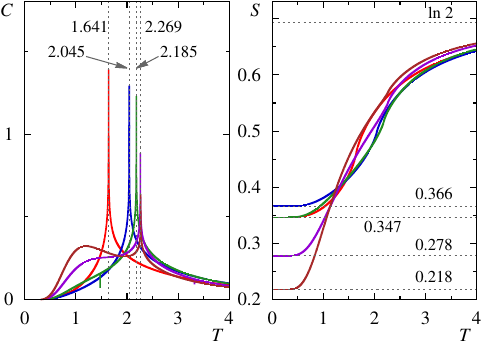}
\caption{Heat capacity (left panel) and entropy (right panel) in semi-decorated square lattice with competing interactions at $d_{y}=0$, $J_{dy}=0$, $J_{y}=+1$, $J_{x}=-1$, and $J_{dx}=+1$. Red curves~-- $d_{x}=1$, blue curves~-- $d_{x}=2$, green curves~-- $d_{x}=3$, violet curves~-- $d_{x}=6$, brown curves~-- $d_{x}=10$}
\label{fig:04}
\end{figure}

\begin{figure}[htb]
\centering \includegraphics{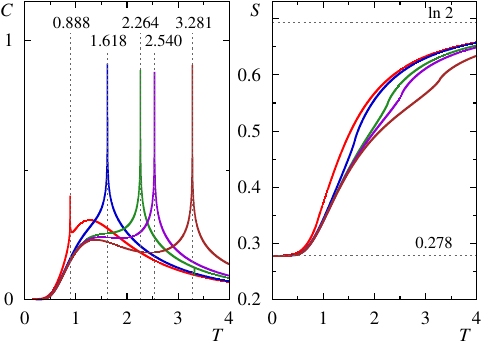}
\caption{Heat capacity (left panel) and entropy (right panel) in semi-decorated square lattice with competing interactions at $d_{x}=6$, $d_{y}=0$, $J_{dy}=0$, $J_{y}=+1$, $J_{x}=-1$, and $J_{dx}=+1$. Red curves~-- $J_{y}=+0.15$, blue curves~-- $J_{y}=+0.5$, green curves~-- $J_{y}=+1$, violet curves~-- $J_{y}=+1.25$, brown curves~-- $J_{y}=+2$}
\label{fig:05}
\end{figure}

As in the previous section~\ref{sec:1r} ``First regime. Free spins in semi-decorated lattice'', for the comprehensive solution of the whole problem, we performed the complete variation of all relevant parameters. The results of a huge array of numerical calculations are depicted to some extent, though not entirely in Figs.~\ref{fig:04} and \ref{fig:05}.

For example, in Fig.~\ref{fig:04} only the multiplicity of decoration $d_{x}$ changes at fixed exchange interactions, but, contrary to Fig.~\ref{fig:02}, in which all the phase-transition points coincide, in the regime with competing direct and decorating exchange interactions both the phase-transition points and residual entropy acquire different values. 

Figure~\ref{fig:05} demonstrates that heat capacity being dependent on only the decoration multiplicity $d_{x}$ has not only different phase-transition points, but also multifarious shapes including those with either nucleating or existing splitting into a sharp peak and a shallow hump, or without splitting at all (see detailed discussion of this phenomenon in Ref.~\citep{Zarubin:2020}). In Figure~\ref{fig:05} it is also seen that the residual entropy keeps one and the same value, as it happens in the regime with free spins (see Fig.~\ref{fig:05}, right panel).

\begin{figure}[htb]
\centering \includegraphics{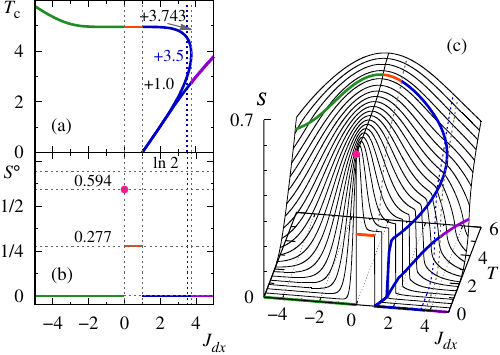}
\caption{Phase-transition temperatures (a) and residual entropy (b) in semi-decorated square lattice with variation of $J_{dx}$ interaction at fixed $d_{x}=6$, $d_{y}=0$, $J_{dy}=0$, $J_{y}=+4$, and $J_{x}=-1$. Lilac circlets and red curve and straight lines~-- ranges of frustrations, blue curves and straight lines~-- non-frustrated three-transition range, green and violet curves and straight lines~-- non-frustrated single-transition ranges. Crossovers from non-frustrated regime to frustrated and reentrantly from frustrated regime to non-frustrated (c)} 
\label{fig:06}
\end{figure}

\begin{figure}[htb]
\centering \includegraphics{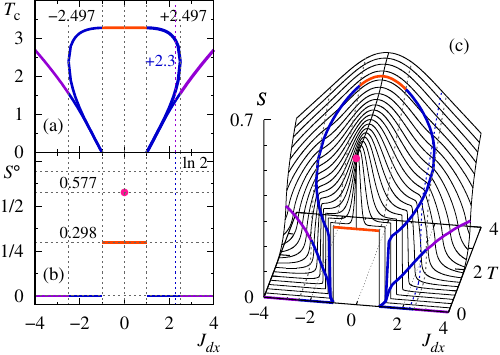}
\caption{Phase-transition temperatures (a) and residual entropy (b) in semi-decorated square lattice with variation of $J_{dx}$ interaction at fixed $d_{x}=5$, $d_{y}=0$, $J_{dy}=0$, $J_{y}=+2$, and $J_{x}=-1$. Lilac circlets and red curve~-- range of frustrations, blue curves~-- non-frustrated three-transition ranges, violet curves and straight lines~-- non-frustrated single-transition ranges. Crossovers from non-frustrated regime to frustrated and reentrantly from frustrated regime to non-frustrated (c)}
\label{fig:07}
\end{figure}

\begin{figure}[htb]
\centering \includegraphics{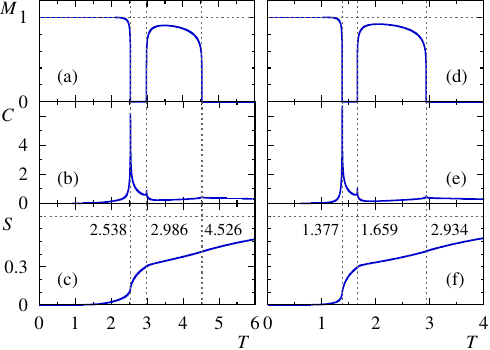}
\caption{Spontaneous magnetization (a, d), heat capacity (b, e), and entropy (c, f) with three phase-transition points at $J_{dx}=+3.5$ for Fig.~\ref{fig:06} (left panel) and at $J_{dx}=+2.3$ for Fig.~\ref{fig:07} (right panel)}
\label{fig:08}
\end{figure}

In Figures~\ref{fig:06} and \ref{fig:07} we set out computations of the phase-transition temperatures and residual entropy with wide variation of $J_{dx}$ decoration exchange interaction, which reveal much more interesting frustration results in comparison with those from section~\ref{sec:1r} ``First regime. Free spins in semi-decorated lattice''. In particular, in the both figures there appear the subranges with simultaneous existence of three phase-transition temperatures denoted by blue color (see Fig.~\ref{fig:08}). But however, no frustrations are observed within all such subranges. From these figures it is seen that the whole range of $J_{dx}$ variation (from $-\infty$ to $+\infty$) subdivides into separate subranges, at the boundaries between which the residual entropy changes jumpwise. The subranges with residual entropy are shown by red. The subranges with zero values of the ground-state entropy are shown by blue and green. The patterns at an even (Fig.~\ref{fig:06}) and at an odd (Fig.~\ref{fig:07}) $d_{x}$ decoration multiplicity differ qualitatively. It should be noted that at fixed exchange interaction with the opposite sign $J_{x}=+1$ Fig.~\ref{fig:06} suffers the mirror reflection with respect to an ordinate axis, while Fig.~\ref{fig:07} with such changing the sign $J_{x}$ from $-1$ to $+1$ suffers no changes at all.

Analyzing Figures~\ref{fig:06} and \ref{fig:07} and comparing them with those from the previous section Figures~\ref{fig:02}--\ref{fig:05} we observe the appearance of some new peculiarities in the behavior of entropy. In the ground state of the system (at $T=0$) there nucleate plateaus at a certain height (red line) and at the bottom (blue and violet lines) and an isolated singular point above (lilac circlet). These plateaus extend over low temperatures and gradually disappear with temperature increasing. The isolated point corresponds to the residual entropy that remains almost unchanged at low temperatures and gradually rising with temperature increasing up to the value of $\ln 2$ at infinite temperature. 

Actually we observe here a new spectacular phenomenon~-- the crossover from one frustration regime (competing) to another (free-spins) or from the frustration regime to non-frustrated regime and even multiple crossovers between various regimes over variation of the exchange interaction rather than conventional crossovers over temperature. Specifically, at the bottom level of Figure~\ref{fig:07}b, a simple crossover occurs~-- from non-frustrated regime to the frustrated one on the left and from the frustrated regime to the non-frustrated one on the right. Above the bottom level of Figure~\ref{fig:07}b, the crossover occurs from the frustrated regime to the same frustrated regime however not directly but across the intermediate regime, which is depicted by lilac circlet and being the inherent feature of the free-spin regime discussed in the previous section.

Keeping all above remarks in mind, we may draw a conclusion that Figs.~\ref{fig:04}--\ref{fig:07} cover all possible frustration properties of semi-decorated square lattice with boundary crossovers to adjacent non-frustrated regimes. All numerical calculations in the regime called semi-decorated square lattice with competing direct interaction and interaction between decorating spins unambiguously show that the residual entropy in all the cases only depends on the decoration multiplicity $d_{x}$ by the following formula
\begin{equation}
S_{\text{comp\ semi}}^{\circ}=\frac{\ln(1+d_{x})}{1+d_{x}},
\label{eq:14}
\end{equation}
with only one exception. It concerns the presence of isolated lilac circlets (residual entropy) in Figs.~\ref{fig:06} and \ref{fig:07} that do not obey the formula~(\ref{eq:14}), but rather the formula~(\ref{eq:13}) from the previous section. This is explained by the fact that throughout the course of $J_{dx}$ interaction variation it vanishes, and, as a consequence, at such points the first mechanism (free-spins) that generates frustrations substitutes the second mechanism (competing). The whole procedure of the exchange interaction variation makes it possible to establish the relationships between various operating mechanisms of frustrations and between various frustration regimes as well.

\section{Third regime. Free spins in both directions of square lattice}
\label{sec:3r}

\begin{figure}[htb]
\centering \includegraphics{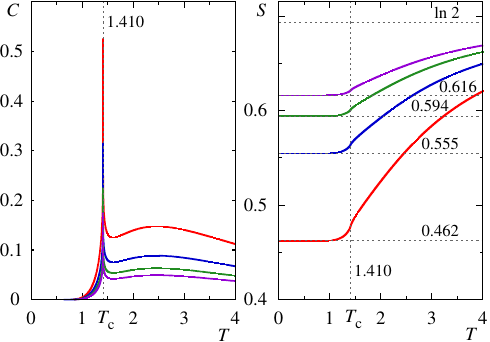}
\caption{Heat capacity (left panel) and entropy (right panel) in square lattice with free spins in both $x$- and $y$-directions at $J_{x}=+3$ and $J_{y}=+0.02$. Red curves~-- $d_{x}=d_{y}=1$, blue curves~-- $d_{x}=d_{y}=2$, green curves~-- $d_{x}=d_{y}=3$, violet curves~-- $d_{x}=d_{y}=4$}
\label{fig:09}
\end{figure}

\begin{figure}[htb]
\centering \includegraphics{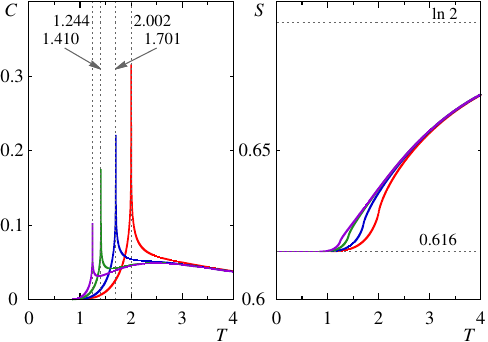}
\caption{Heat capacity (left panel) and entropy (right panel) in square lattice with free spins in both $x$- and $y$-directions at $J_{x}=+3$. Red curves~-- $d_{x}=8$, $d_{y}=0$, $J_{y}=+0.1$; blue curves~-- $d_{x}=7$, $d_{y}=1$, $J_{y}=+0.05$; green~-- curves $d_{x}=6$, $d_{y}=2$, $J_{y}=+0.02$; violet curves~-- $d_{x}=4$, $d_{y}=4$, $J_{y}=+0.01$}
\label{fig:10}
\end{figure}

Let us consider now a special case when in both directions of the square lattice there operates the first mechanism (free-spins) generating frustrations. The vast computation arrays are exemplified by Figures~\ref{fig:09} and \ref{fig:10}.

In Figure~\ref{fig:09} at certain fixed values of direct exchange interactions, namely, $J_{x}=+3$ and $J_{y}=+0.02$, while the both multiplicities $d_{x}$ and $d_{y}$ vary simultaneously being equal to one another.

In Figure~\ref{fig:10} contrary to Figure~\ref{fig:09} the direct exchange interaction $J_{y}$ vary at the direct exchange interaction $J_{x}$ being kept constant (equal to~$3$), while the multiplicities $d_{x}$ and $d_{y}$ vary such that their sum does not change at all and equals $d_{x}+d_{y}=8$.

All numerical calculations in the regime called free spins in both directions of decorated square lattice unambiguously show that the residual entropy in all the cases only depends on the decoration multiplicities $d_{x}$ and $d_{y}$ by the following formula
\[
S_{\text{free-free}}^{\circ}=\frac{(d_{x}+d_{y})\ln2}{1+d_{x}+d_{y}}.
\]
The loci of the phase-transition temperatures depend only on the direct exchange interactions $J_{x}$ and $J_{y}$, which corresponds to non-frustrated square lattice, and they are determined entirely accordant to Onsager's formulas~\citep{Onsager:1944}, which allow only one phase transition. Therefore, contrary to the previous regime ``Free spins in semi-decorated lattice'' (section~\ref{sec:1r}), in which simultaneous existence of three phase-transition temperatures were observed, in the discussed regime ``Free spins in both directions of square lattice'' not a single sign of multiple transitions has been revealed at all.

\section{Fourth regime. Free spins in one direction of square lattice}
\label{sec:4r}

In this section we will discuss the regime in which in one direction, namely, ($x$-direction without loss of generality) of the square lattice the free-spins operates as the mechanism of creating frustrations, which means that the direct exchange interaction $J_{x}$ may be both either positive (ferromagnetic) or negative (antiferromagnetic) and may have arbitrary values excluding zero, while the decorating exchange interaction $J_{dx}$ should be equal to zero.

In the other direction, the requirements imposed on the direct interaction $J_{y}$ and the interaction between decorating spins and between the nodal and decorating spins $J_{dy}$ should thoroughly satisfy the following conditions. First of all, they should not be equal to zero. Second, they may simultaneously be either positive (ferromagnetic) or negative (antiferromagnetic). The last not least, even if these interactions ($J_{y}$ and $J_{dy}$) turn out to be competing, which depend on the interactions signs and the parities of decoration multiplicity $d_{y}$, at odd multiplicity $d_{y}$ the interaction $J_{y}$ must be only negative (antiferromagnetic) and the corresponding interaction $J_{dy}$ may be both negative (antiferromagnetic) and positive (ferromagnetic). In the opposite case, namely, at even multiplicity $d_{y}$ the interaction $J_{y}$ may be negative and the corresponding interaction $J_{dy}$ must be positive, or contrariwise, $J_{y}$ positive and $J_{dy}$ negative. In any case, the values of competing interactions $J_{y}$ and $J_{dy}$ should comply with the strict requirement 
\[
|J_{y}|<|J_{dy}|,
\]
that is actually an opposite condition to the requirement for the appearance of frustrations, see Eq.~(\ref{eq:reqx}).

\begin{figure}[htb]
\centering \includegraphics{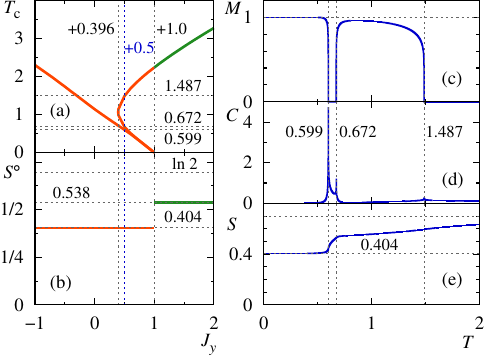}
\caption{Phase-transition temperatures (red and green curves) (a) and residual entropy (red and green straight lines) (b) in decorated ($d_{x}=7$, $d_{y}=4$) square lattice with free spins in $x$-direction at $J_{x}=-1$, $J_{dx}=0$, $J_{dy}=-1$, and variation of direct exchange interaction $J_{y}$. Spontaneous magnetization (c), heat capacity (d), and entropy (e) with three phase-transition points at $J_{y}=+0.5$}
\label{fig:11}
\end{figure}

In the left panel of Figure~\ref{fig:11} we present the computation results of the phase diagram (phase-transition temperatures) of decorated (multiplicities $d_{x}=7$ and $d_{y}=4$) square lattice in the free-spin regime along $x$-direction with three fixed exchange interaction parameters: $J_{x}=-1$, $J_{dx}=0$, $J_{dy}=-1$. In this diagram there have been calculated the phase-transition temperature and entropy as the functions of the direct exchange $J_{y}$ that varied in a wide range of negative and positive values. It turned out that the whole range separates into three subranges, such that in the middle subrange the system has three phase-transition temperatures while in the left-hand and right-hand semi-infinite subranges one and only phase-transition temperature.

The right panel of Figure~\ref{fig:11} demonstrates the calculation of the spontaneous magnetization, heat capacity, and entropy at a separate point $J_{y}=+0.5$ (denoted in the left panel of Fig.~\ref{fig:11} by a vertical blue dashed line) that shows the so-called reentrance effect (see Ref.~\citep{Diep:2020}). It should be specially pointed out here that in Fig.~\ref{fig:11} a new spectacular phenomenon~-- the coexistence of the frustrations and reentrancy is observed, which additionally refutes the widespread myth about frustrations and phase transitions as mutually exclusive phenomena.

\begin{figure}[htb]
\centering \includegraphics{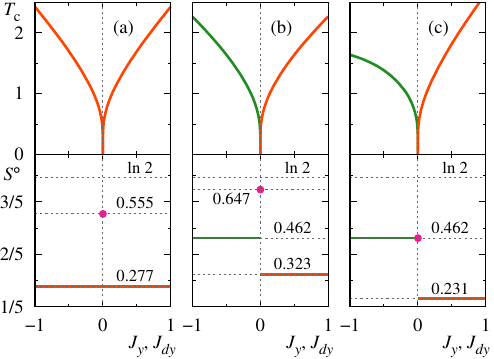}
\caption{Phase-transition temperatures (upper panels) and residual entropies (lower panels) in square lattice with free spins in $x$-direction at $J_{x}=-1$, $J_{dx}=0$, and simultaneous variations of direct exchange interaction $J_{y}$ and decorating exchange interaction at $J_{y}=J_{dy}$. Multiplicities $d_{x}=d_{y}=2$ (a), $d_{x}=d_{y}=7$ (b), and $d_{x}=d_{y}=1$ (c). Red lines correspond to the regime when in $y$-direction there are neither free spins, nor competition between direct and decorating interaction. Lilac circlets correspond to the regime with free spins in both $x$- and $y$-directions. Green lines correspond to the regime with free spins in $x$-direction and competition of direct and decorating interaction in $y$-direction}
\label{fig:12}
\end{figure}

For the purpose of more detailed investigation of frustration properties of decorated square lattice in the free-spins regime along one direction (with $J_{x}=-1$ and $J_{dx}=0$) we have undertaken an improved technique of simultaneous varying two exchange interactions in the other direction, direct $J_{y}$ and decorating $J_{dy}$. The Figure~\ref{fig:12} presents an example of this technique.

\begin{figure}[htb]
\centering \includegraphics{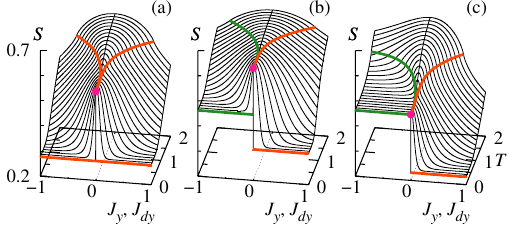}
\caption{Crossovers in square lattice with free spins in $x$-direction at $J_{x}=-1$, $J_{dx}=0$, and simultaneous variations of direct exchange interaction $J_{y}$ and decorating exchange interaction at $J_{y}=J_{dy}$. Multiplicities $d_{x}=d_{y}=2$ (a), $d_{x}=d_{y}=7$ (b), and $d_{x}=d_{y}=1$ (c)}
\label{fig:13}
\end{figure}

The upper panels of Figure~\ref{fig:12} present the results of calculating phase-transition-temperature, while the lower panels of Figure~\ref{fig:12} demonstrate the calculated residual entropy at these parameters. Figure~\ref{fig:13} shows the temperature dependence of entropy.

The left panel of Figures~\ref{fig:12}a and~\ref{fig:13}a is represented by even values of multiplicities ($d_{x}=d_{y}=2$), from which it can be seen that with negative values of the exchange interactions $J_{y}$ and $J_{dy}$ in the ground state of the system (at $T=0$), residual entropy arises at a level equal to $\approx {0.277}$. Then, with increasing temperature, it almost does not change, forming a kind of plateau. In the process of varying $J_{y}$ and $J_{dy}$ , when they reach zero values, the residual entropy increases by a leap, taking a value equal to $\approx {0.555}$, and with further variation it decreases by a leap, returning to the previous level equal to $\approx {0.277}$. Thus, peculiar crossovers of the residual entropy are observed from one frustrating regime to an intermediate one and subsequent return to the original one (that is, reentrantly). At elevated temperatures, both plateaus disappear smoothly.

The middle panel of Figures~\ref{fig:12}b and~\ref{fig:13}b is represented by odd frustration multiplicities $d_{x}=d_{y}=7$, which shows that the whole process resembles the previous one. However, there are significant differences. Firstly, all the heights of the plateau are different and different from the previous ones. Secondly, crossovers occur from one frustrating regime to another and then to a third without returning to the original one.
 
The right panel of Figures~\ref{fig:12}c anf~\ref{fig:13}c is characterized by odd multiplicities $d_{x}=d_{y}=1$, and the whole process also resembles the previous one. The differences are such that the heights of the plateau are different, but at these multiplicities the residual entropies of the intermediate and right ranges coincided (due to an accidental coincidence of expressions for residual entropy, corresponding to the regime with free spins in both $x$- and $y$-directions and that which corresponds to the regime with free spins in one direction and the regime with competition of direct and decorating interaction in the other direction).

All numerical calculations in the regime considered in the present section unambiguously show that the residual entropy in all the cases only depends on the decoration multiplicities $d_{x}$ and $d_{y}$ by the following equivalent formulas:
\begin{align*}
S_{\text{free-nc}}^{\circ}&=\frac{d_{x}\ln2}{1+d_{x}+d_{y}}, \\
S_{\text{nc-free}}^{\circ}&=\frac{d_{y}\ln2}{1+d_{x}+d_{y}}.
\end{align*}

These formulas and the assembled debates make it possible to establish the relationships between the frustration regimes from the present section and the previous and the following ones.

\section{Fifth regime. Competing interactions in square lattice}
\label{sec:5r}

This section is aimed to investigate the regime in which in one direction ($x$-direction without loss of generality) of the square lattice the competition between the direct and decorating exchange interaction plays the role of the mechanism that create frustrations, which means that in one choice (with odd multiplicity $d_{x}$ the direct exchange interaction $J_{x}$ may only be negative (antiferromagnetic) and the decorating interaction $J_{dx}$ either positive (ferromagnetic) or negative (antiferromagnetic) and in another choice (with even multiplicity $d_{x}$) the direct exchange interaction $J_{x}$ may both be positive (ferromagnetic) or negative (antiferromagnetic). In this choice the decorating interaction $J_{dx}$ should have opposite sign to $J_{x}$, namely, negative (antiferromagnetic) or positive (ferromagnetic). In both the choices the values of $J_{x}$ and $J_{dx}$ have to meet the necessary requirement for the appearance of frustrations (Eq.~(\ref{eq:reqx})).

In the other direction the requirements imposed on the direct interaction $J_{y}$ and the interaction between decorating spins and between the nodal and decorating spins $J_{dy}$ should thoroughly satisfy the following conditions. 

First of all, $J_{dy}$ should not be equal to zero. Second, they may simultaneously be either positive (ferromagnetic) or negative (antiferromagnetic). The last not least, even if these interactions ($J_{y}$ and $J_{dy}$) turn out to be competing, which depend on the interactions signs and the parities of decoration multiplicities ($d_{x}$ and $d_{y}$), then at odd multiplicity $d_{y}$ the interaction $J_{y}$ must be only positive (ferromagnetic) and the corresponding interaction $J_{dy}$ may both be negative (antiferromagnetic) and positive (ferromagnetic), while at even multiplicity $d_{y}$ the interaction $J_{y}$ and the corresponding interaction $J_{dy}$ ought to be simultaneously either positive (ferromagnetic) or negative (antiferromagnetic).

\begin{figure}[htb]
\centering \includegraphics{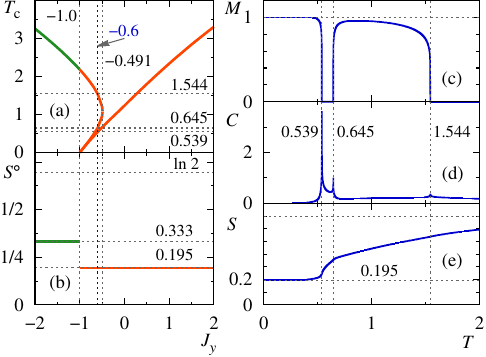}
\caption{Phase-transitions temperatures (red and green curves) (a) and residual entropies (red line, green straight line) (b) in decorated ($d_{x}=6$, $d_{y}=3$) square lattice with competing direct interaction $J_{x}=-1$ and decorating interaction $J_{dx}=+0.7$ at $J_{dy}=+1$ as the functions of $J_{y}$. Spontaneous magnetization (c), heat capacity (d), and entropy (e) with three phase-transition points at $J_{y}=-0.6$}
\label{fig:14}
\end{figure}

In the left panel of Figure~\ref{fig:14}a we present the phase diagram of decorated (multiplicities $d_{x}=6$ and $d_{y}=3$) square lattice in current section ``Fifth regime. Competing interactions in square lattice'' along $x$-direction with three fixed exchange interaction parameters: direct interaction $J_{x}=-1$, decorating interaction $J_{dx}=+0.7$ and $J_{dy}=+1$ with variation of the direct exchange interaction $J_{y}$.

The right panel of Figure~\ref{fig:14} shows the temperature dependences of the spontaneous magnetization, heat capacity, and entropy of the system at the chosen fixed value of $J_{y}=-0.6$ denoted in the left panel of Figure~\ref{fig:14} by a vertical blue dashed line. From this figure it is seen that the non-zero residual entropy may be observed not only in one-phase-transition case, but even in the three-phase-transition one. The Figure~\ref{fig:14} demonstrates also the so-called reentrance effect and a simple crossover to the regime of competing interactions in the both directions of decorated square lattice.

\begin{figure}[htb]
\centering \includegraphics{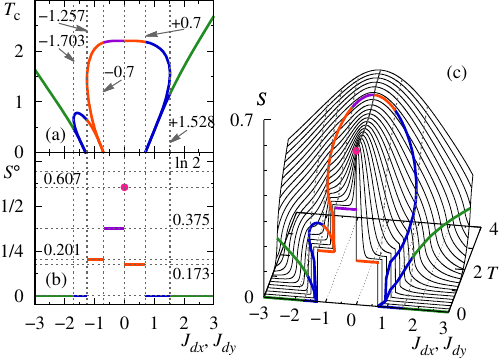}
\caption{Phase-transition temperatures (red, blue, and green curves) (a) and residual entropies (green, blue, red, violet straight lines and lilac circlet) (b) in decorated ($d_{x}=3$, $d_{y}=4$) square lattice with competing interactions. The direct interaction $J_{x}=-0.7$ and direct interaction $J_{y}=+1.3$. Five crossovers with the three three-phase-transition-temperature ranges (c). Variation is taken simultaneously over both $J_{dx}=J_{dy}$. The two subranges denoted only by red correspond to ``Competing interactions in square lattice'' regime}
\label{fig:15}
\end{figure}

In the left panel of Figure~\ref{fig:15} there are presented the phase-transition temperatures and residual entropies of decorated square lattice with multiplicities $d_{x}=3$ and $d_{y}=4$ in the fifth regime ``Competing interactions in square lattice'' with two fixed direct exchange interactions $J_{x}=-0.7$ and $J_{y}=+1.3$ in the course of simultaneous variation of the both direct exchange interactions $J_{dx}$ and $J_{dy}$.

Simple non-frustrated ranges with one phase transition are denoted by green, the non-frustrated ranges with three phase transitions are denoted by blue. Two frustrated ranges with one or three phase transitions are denoted by red, and they correspond to the fifth regime that is discussed in the current section, but however they have different values of residual entropy. There are also observed two frustrated ranges with one phase transition, one of which (denoted by violet) is attributed to the regime with both competing interactions, and another one (denoted by lilac circlet) is attributed to the both free-spins regime. 

The right panel of Figure~\ref{fig:15} presents the temperature evolution of the phase-transition temperature and entropy, and also five crossovers between plateaus that nucleate in the ground state of the system and fade away at elevated temperatures.

On the whole, the Figure~\ref{fig:15} represents rather complicated pattern of sequentially alternating eight subranges of the $J_{dx}$ and $J_{dy}$ variation with various types of physical properties and five crossovers between various residual-entropy-driven regimes.

\begin{figure}[htb]
\centering \includegraphics{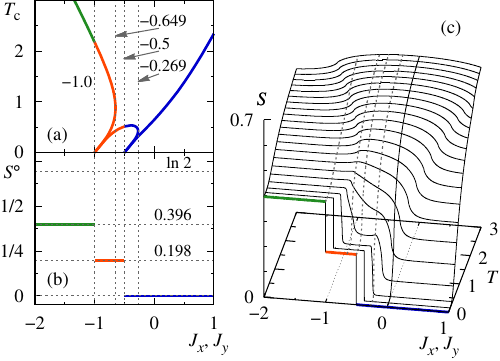}
\caption{Phase-transition temperatures (green, red, and blue curves) (a) and residual entropies (green, red, and blue straight lines) (b), in decorated ($d_{x}=3$, $d_{y}=3$) square lattice with competing interactions. The decorating exchange interaction $J_{dx}=-1$ and decorating interaction $J_{dy}=-0.5$. Variation is taken simultaneously over both direct exchange interactions $J_{x}=J_{y}$. Three crossovers with the three three-phase-transition-temperature ranges (c). The subranges denoted only by red correspond to regime ``Fifth regime. Competing interactions in square lattice''} 
\label{fig:16}
\end{figure}

The Figures~\ref{fig:15} and \ref{fig:16} demonstrate the distinctive feature of the fifth regime ``Competing interactions in square lattice'', that is the existence of two and even three three-phase-transition-temperature subranges. For perception convenience, all the subranges are separated by vertical dashed lines that simultaneously denote the boundaries between adjacent regimes with other residual entropies including those without frustrations at all.

All numerical calculations in the regime considered in the present section unambiguously show that the residual entropy in all the cases only depends on the decoration multiplicities $d_{x}$ and $d_{y}$ by the following equivalent formulas:
\begin{align*}
S_{\text{comp-nc}}^{\circ}&=\frac{\ln(1+d_{x})}{1+d_{x}+d_{y}}, \\
S_{\text{nc-comp}}^{\circ}&=\frac{\ln(1+d_{y})}{1+d_{x}+d_{y}}.
\end{align*}

\section{Sixth regime. Competing interactions and free spins in square lattice}
\label{sec:6r}

In this section we study the regime in which the requirements imposed on all the parameters in one direction are very similar to those from the previous section. They are such as: in $x$-direction of the square lattice the competition between the direct and decorating exchange interaction plays the role of the mechanism that create frustrations, which means that in one choice (with odd multiplicity $d_{x}$ the direct exchange interaction $J_{x}$ may only be negative (antiferromagnetic) and the decorating interaction $J_{dx}$ either positive (ferromagnetic) or negative (antiferromagnetic) and in another choice (with even multiplicity $d_{x}$) the direct exchange interaction $J_{x}$ may both be positive (ferromagnetic) or negative (antiferromagnetic). In this choice the decorating interaction $J_{dx}$ should have opposite sign to $J_{x}$, namely, negative (antiferromagnetic) or positive (ferromagnetic). In both the choices the values of $J_{x}$ and $J_{dx}$ have to meet the necessary requirement for the appearance of frustrations (Eq.~(\ref{eq:reqx})).

In the other direction the requirements are very simple. The decorating interaction $J_{dy}$ should uniquely be equal to zero, while contrariwise, the direct interaction $J_{y}$ ought not to be equal to zero. 

\begin{figure}[htb]
\centering \includegraphics{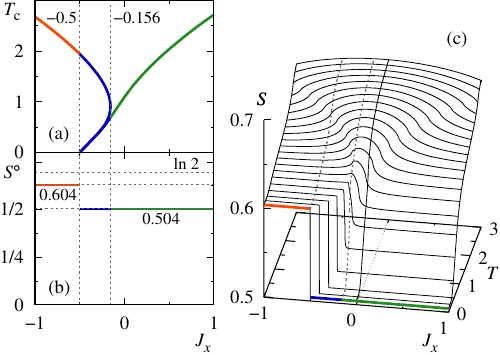}
\caption{Phase-transition temperatures (red, blue, and green curves) (a) and residual entropies (red, blue, and green straight lines) (b) in decorated ($d_{x}=2$, $d_{y}=8$) square lattice in the sixth regime ``Competing interactions and free spins in square lattice'' with competing interactions $J_{x}$ and $J_{dx}=+0.5$ at $J_{y}=-1.4$ and $J_{dy}=0$. Variation is taken over the direct exchange interaction $J_{x}$. Simple crossover from frustrated regime to non-frustrated regime with a saltus (c)}
\label{fig:17}
\end{figure}

For a profound understanding of the frustrating processes occurring on a decorated square lattice in the regime considered in this section, we present here two completely different figures. Let us consider firstly Fig.~\ref{fig:17} and compare it with Fig.~\ref{fig:14} from the previous section.

The left panel of Figure~\ref{fig:17} shows the dependences of the phase-transition temperature and residual entropy in the course of continuous variation of the direct exchange interaction $J_{x}$ from negative to positive values. The entire calculated diagram splits into three subranges: one-phase-transition subrange to three-phase-transition one and again a return to one-phase-transition subrange. At the same time, entropy experiences only one saltus with a decrease in magnitude. It is very wondering that the whole considered process exactly resembles that shown in the left panel of Figure~\ref{fig:14}. Actually, these figures are topologically equivalent, which leads to an attractive interpretation of physical phenomena equivalence in both cases. However, there is a significant difference between them.

In Figure~\ref{fig:17} the nucleating residual entropy is attributed to the sixth regime ``Competing interactions and free spins in square lattice'' and then suffers a crossover to the fifth regime ``Competing interactions in square lattice'' (section~\ref{sec:5r}), while in Figure~\ref{fig:14} residual entropy nucleates with a value that corresponds to the sequential seventh regime ``Competing interactions in both directions of square lattice'' (section~\ref{sec:7r}). Then it returns to a value attributing to the fifth regime ``Competing interactions in square lattice'' (section~\ref{sec:5r}).

The right panel of Figure~\ref{fig:17}c demonstrates a simple crossover between two different frustrated regimes.

\begin{figure}[htb]
\centering \includegraphics{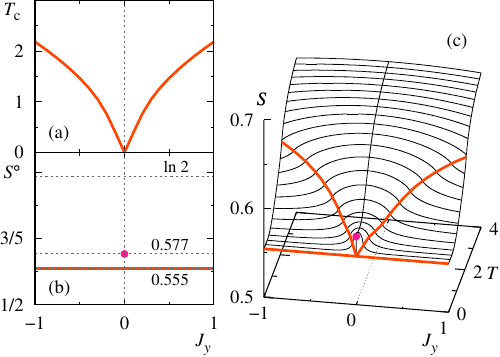}
\caption{Phase-transition temperatures (red curves) (a) and residual entropies (red straight line and lilac circlet) (b) in decorated ($d_{x}=3$, $d_{y}=6$) square lattice in the sixth regime ``Competing interactions and free spins in square lattice'' at direct interaction $J_{x}=-1$, decorating interaction $J_{dx}=+1$ and $J_{dy}=0$. Variation is taken over direct exchange interaction $J_{y}$. Crossover from frustrated regime reentrantly to the same frustrated regime across a singular point (c)}
\label{fig:18}
\end{figure}

Figure~\ref{fig:18} shows prima facie a simple reenterable crossover between plateaus with equal entropies across a singular point (with another residual entropy $\approx0.576\,832$) much similar to that in the left panel of Figure~\ref{fig:12}, which seems that it is nothing to argue here. However, it is just this numerical value of residual entropy at a singular point that turned out to be a trigger point to a challenge. First of all, it appeared that this value was not found in all of the frustration regimes, offered in our paper. Moreover, the detailed analysis has shown that residual entropies with such a property were observed in a wide variety of the arrays with different multiplicities $d_{x}$ and $d_{x}$ and also at many exchange interactions $J_{x}$ and $J_{dx}$ that obey the necessary requirement for the appearance of frustrations (Eq.~(\ref{eq:reqx})). We were even lucky to derive the most general formula for such residual entropies
\begin{multline}
S^{\circ}=\ln2+\frac{1}{4\pi(1+d_{x}+d_{y})}\intop_{0}^{2\pi}\ln\Biggl[\frac{d_{x}^{2}+2d_{x}+2}{2^{2d_{x}+1}}\\
+\frac{d_{x}(d_{x}+2)}{2^{2d_{x}+1}}\cos\phi\Biggr]\,d\phi.
\label{eq:16:}
\end{multline}

It is worthy to specially be noticed that this formula is applicable not to all possible cases, but only for the exchange interactions $J_{x}$ and $J_{dx}$ that comply the narrowed requirement for the appearance of frustrations $|J_{x}|=|J_{dx}|$.

\begin{figure}[htb]
\centering \includegraphics{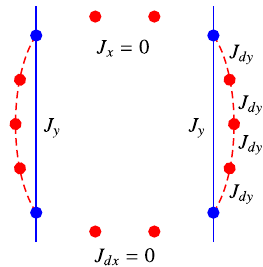}
\caption{Square lattice with free spins unbound in one direction both by direct $J_x$ and decorating $J_{dx}$ exchange interactions}
\label{fig:19}
\end{figure}

We have also found an acceptable solution to the challenge. In the whole our paper under the abbreviation ``free spins'' it was implied that free spins are unbound by only the decorating exchange interaction (say $J_{dx}$), while in the same direction of the square lattice the nodal spins are nevertheless bound by the direct interaction $J_{x}$. As opposed to above implied concept of ``free spins'' the square lattice with the residual-entropy value $\approx0.576\,832$ was obtained at $J_{x}=0$ and $J_{dx}=0$, which means that here \textit{all the spins both decorating and nodal in this direction are unbound}. Eventually, the obtained system by no means should be called ``2D square lattice'', but rather an array of self-dependent 1D chains with unbound spins in between by and atour (see Fig.~\ref{fig:19}).

The numerical calculations in the regime considered in the present section unambiguously show that the residual entropy in all the cases only depends on the decoration multiplicities $d_{x}$ and $d_{y}$ by the following equivalent formulas:
\begin{align*}
S_{\text{comp-free}}^{\circ}&=\frac{\ln(1+d_{x})+d_{y}\ln2}{1+d_{x}+d_{y}}, \\
S_{\text{free-comp}}^{\circ}&=\frac{d_{x}\ln2+\ln(1+d_{y})}{1+d_{x}+d_{y}}.
\end{align*}

\section{Seventh regime. Competing interactions in both directions of square lattice}
\label{sec:7r}

The ultimate goal of all our investigations is the most intricate and therefore catching and wealthy in details problem~-- the regime with competing interactions in the square lattice multiply decorated in both directions.

The requirements that are imposed on all the parameters involved in studying are just the same for both lattice directions, being nevertheless different in values. Specifically they depend on the relationships between the signs and values of exchange interactions ($J_{x}$, $J_{y}$, $J_{dx}$ $J_{dy}$), as well as on the parities of decoration multiplicities ($d_{x}$ $d_{y}$).

The role of the mechanisms that create frustrations is played by the competition between the direct and decorating exchange interactions including two variant of choices: at odd multiplicity $d_{x}$ ($d_{y}$) the direct exchange interaction $J_{x}$ ($J_{y}$) may only be negative (antiferromagnetic) and the decorating interaction $J_{dx}$ ($J_{dy}$) may be either positive (ferromagnetic) or negative (antiferromagnetic) and in another variant (at even multiplicity $d_{x}$ ($d_{y}$)) the direct exchange interaction $J_{x}$ ($J_{x}$) may both be positive (ferromagnetic) or negative (antiferromagnetic). In this variant the decorating interaction $J_{dx}$ ($J_{dy}$) ought to have opposite sign to $J_{x}$ ($J_{y}$), namely, negative (antiferromagnetic) or positive (ferromagnetic). In both the variants, the values of $J_{x}$ ($J_{x}$) and $J_{dx}$ ($J_{dy}$) must obey the necessary requirement for the appearance of frustrations (Eq.~(\ref{eq:reqx})).

\begin{figure}[htb]
\centering \includegraphics{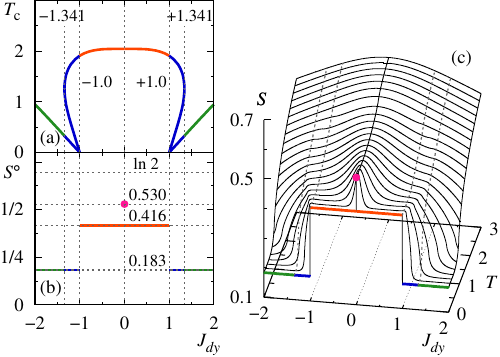}
\caption{Phase-transition temperatures (green, blue, and red curves) (a) and residual entropies (green, blue, and red straight lines and lilac circlet) (b) in decorated ($d_{x}=2$, $d_{y}=3$) square lattice in the seventh regime ``Competing interactions in both direction of square lattice'' at direct interaction $J_{x}=-1$, direct interaction $J_{y}=-1$, decorating interaction $J_{dx}=+1$. Variation is taken over decorating exchange interaction $J_{dy}$. Four crossovers across a singular point with reentrance effect (c)}
\label{fig:20}
\end{figure}

We presented in Figure~\ref{fig:20} one of many possible variants of multiple crossovers between various frustration states, which are frequently met in the seventh regime ``Competing interactions in both directions of square lattice''. To this end, we have chosen the following set of fixed parameters: direct interaction $J_{x}=-1$, direct interaction $J_{y}=+1$, decorating interaction $J_{dx}=+1$, and varying decorating exchange interaction $J_{dy}$ from negative to positive values.

The left panel of Figure~\ref{fig:20} shows the computations of the phase diagram of the system~-- phase-transition temperatures and residual entropy. The temperature dependence of entropy and crossover effects are shown in the right panel of Figure~\ref{fig:20}c, in particular across a singular point, the residual entropy of which belongs to the fifth regime ``Competing interactions in square lattice'' (section~\ref{sec:5r}).

In the issue, we may see four crossovers with two reentrancies.
It is worthy to notice that the phase-transition temperatures and entropy, which are intrinsic to the seventh regime, are shown by red in Fig.~\ref{fig:20}.

\begin{figure}[htb]
\centering \includegraphics{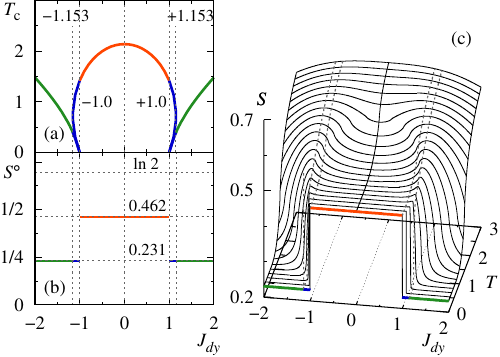}
\caption{Phase-transition temperatures (red, blue, and green curves) (a) and residual entropies (red, blue, and green lines) (b) decorated square lattice with competing interactions in both directions: $J_{x}=-1$ and $J_{dx}=+0.5$ and $J_{y}=-1$ as the functions of $J_{dy}$. Multiplicities $d_{x}=d_{y}=1$. Crossover from frustrated regime reentrantly to the same frustrated regime across another frustrated regime (c)}
\label{fig:21}
\end{figure}

Figure~\ref{fig:21} is another variant of multiple crossovers between the seventh regime ``Competing interactions in both directions of square lattice'' and adjacent frustration states. This figure has much in common with the previous Figure~\ref{fig:20}. In particular, the first and the last crossovers occur specifically from the sixth regime. ``Competing interactions and free spins in square lattice'' (section~\ref{sec:6r}) to the seventh regime, and inversely, from the seventh to the six regime. However, Figure~\ref{fig:21} differs from Figure~\ref{fig:20} by significant attributes, namely, by the absence of a singular point and as a consequence, by the absence of two crossovers. Such a discrepancy looks a little bit strange. Actually, the singular point is also present in Fig.~\ref{fig:21} but it is invisible and does not produce such effect as in Fig.~\ref{fig:20} since at the particular multiplicities $d_{x}=1$ and $d_{y}=1$ the residual entropies from the seventh and fifth regimes suffer a chance coincidence.

The numerical calculations in the regime considered in the present section unambiguously show that the residual entropy in all the cases only depends on the decoration multiplicities $d_{x}$ and $d_{y}$ by the following formula:
\begin{multline*}
S_{\text{comp-comp}}^{\circ}=\\
\frac{1}{2\pi(1+d_{x}+d_{y})}\intop_{0}^{\pi}\ln\frac{1}{2}\Biggl[(d_{x}^{2}+2d_{x}+2)(d_{y}^{2}+2d_{y}+2)\\
-2(d_{x}+1)d_{y}(d_{y}+2)\cos\phi
+\Bigl\{\bigl[(d_{x}^{2}+2d_{x}+2)(d_{y}^{2}+2d_{y}+2)\\
-2(d_{x}+1)d_{y}(d_{y}+2)\cos\phi\bigr]^{2}
-4d_{x}^{2}(d_{x}+2)^{2}(d_{y}+1)^{2}\Bigr\}^{1/2}\Biggr]\,d\phi.
\end{multline*}

The next Figure~\ref{fig:22} shows one more example of multiple crossovers in the seventh regime ``Competing interactions in both direction of square lattice'' featuring the stepwise sequence of crossovers from the fifth frustrated regime across a singular point (appropriate to the sixth regime) to the seventh regime and returning again reentrantly to the fifth frustrated regime.

\begin{figure}[htb]
\centering \includegraphics{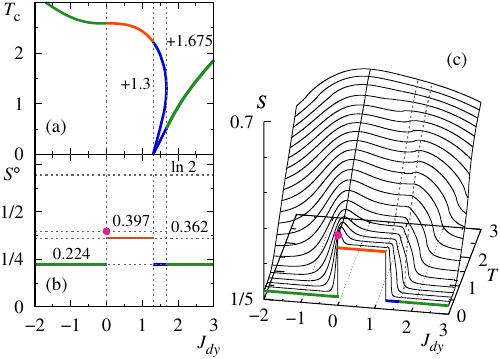}
\caption{Phase-transition temperatures (green, red, and blue curves) (a) and residual entropies (green, red, and blue straight lines) (b) in decorated square lattice with competing interactions in both directions: $J_{x}=-1$ and $J_{dx}=+1$ and direct interaction $J_{y}=-1.3$. Multiplicities $d_{x}=5$ and $d_{y}=2$. Variation is taken over decorating exchange interactions $J_{dy}$. Three crossovers from the fifth regime to the seventh regime and come back to the fifth regime (c)} 
\label{fig:22}
\end{figure}

\begin{figure}[htb]
\centering \includegraphics{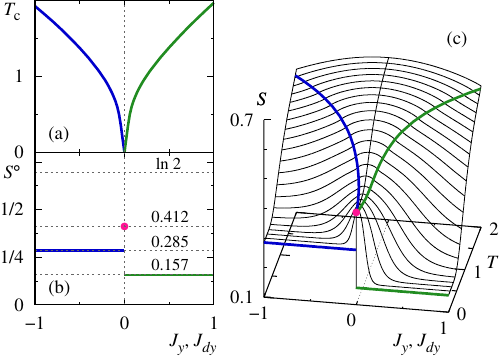}
\caption{Phase-transition temperatures (a) and residual entropies (b) in decorated square lattice in the regime with competing interactions in both directions: $J_{x}=-0.73$ and $J_{dx}=+0.73$ as the function of simultaneous variation of the direct and decorating exchange interactions $J_{y}=J_{dy}$. Multiplicities $d_{x}=8$ and $d_{y}=5$. Double-crossover between two frustrated regimes (from the seventh to the fifth) across singular point that is inappropriate for any of the seven frustration regimes decorated square lattice (c)}
\label{fig:23}
\end{figure}

We also brought about in the present section an extra Figure~\ref{fig:23} with characteristic feature similar to Figure~\ref{fig:18} that has been discussed in detail in the section~\ref{sec:7r}, concerning the reenterable crossover between plateaus with equal entropies across a singular entropy-point, which is inappropriate for any of the seven frustration regimes. Despite the parameters and the value of entropy-point, which appear in Fig.~\ref{fig:23}, differ from those in Fig.~\ref{fig:18}, the newly obtained value corresponds exactly to Eq.~(\ref{eq:16:}). All this arguing on the one hand gives evidence to the generality of the observed effect, and on the other hand to the validity of formula~(\ref{eq:16:}), obtained in the section~\ref{sec:7r}.

Special attention should be paid to the formula derived in this section, ``Competing interactions in both direction of square lattice'' due to the fact that it differs significantly from all others in both size and form, representing Riemann integral that does not reduce to a simple algebraic form, as expressions for residual entropy in all other six regimes. The derivation of this formula was a formidable challenge and required a lot of time, as well as stubborn efforts.

\section{Conclusions}

In the present paper the Ising model on a decorated square lattice with arbitrary multiplicities of decorating spins on horizontal and vertical bonds of the lattice, respectively, is examined within an exact analytical approach based on the solution of the Kramers--Wannier transfer-matrix.

It is established that the frustrations may be triggered by only two mechanisms acting independently in both directions of multiply decorated square lattice.

The first mechanism (called ``free-spins mechanism'') operates when in one and the same direction the lattice has some number of decorating spins, but they do not interact with one another keeping nevertheless direct interaction between the nodal spins.

The second mechanism (called ``competing interactions mechanism'') operates in the presence of direct exchange interaction between nodal spins and the interactions between decorating spins and between decorating and nodal spins. The competing interactions to create frustrations should subject to the following conditions depending on the interactions signs and the parities of decoration multiplicities ($d_{x}$ and $d_{y}$). At odd multiplicity $d_{x}$ the interaction $J_{x}$ must be only negative (antiferromagnetic) and the corresponding interaction $J_{dx}$ may be both negative (antiferromagnetic) and positive (ferromagnetic). At even multiplicity $d_{x}$ the interaction $J_{x}$ may be negative and the corresponding interaction $J_{dx}$ must be positive, or contrariwise, $J_{x}$ positive and $J_{dx}$ negative.

The third mechanism that is usually but erroneously called ``geometrical frustrations'' actually means the interference between antiferromagnetic exchange interactions that couple spins in all three directions of a lattice, which originated in case of the triangular \citep{Wannier:1950,Wannier:1973} or kagome \citep{Kano:1953} lattices. However, as for the square lattice with two orthogonal directions, such an interference is absent, and therefore the third mechanism does not operate in it.

The various combinations of two mechanisms determine an array of seven possible regimes of frustrations, each of which is described by appropriate analytical expression for the residual entropy.

The specific features of every regime and the general properties intrinsic to all regimes are revealed. In particular, contrary to the widespread opinion that the frustration and phase transition are mutually exclusive phenomena, the coexistence of frustrations and the phase transition and even the multiple transitions turned out to be rather a rule but not the exception.

It is established that the found seven regimes are by no means isolated, but nevertheless possess certain relationships between adjacent regimes that are expressed by specific crossovers over the exchange-interaction variation rather then by conventional crossovers over temperature. Such crossovers may be of various types either direct or via intermediate regimes with or without reentrancy. In all seven regimes frustrations give rise to several entropy plateaus that nucleate in the ground state of the system, fade away at elevating temperature, and then form the unified plateau at infinite temperatures at a level of $\ln2$, which is the immanent property of the classical Ising model with two spin states irrespective of the number and exchange-interaction forms.

All the newly observed phenomena are obtained by specially developed qualitative technique and detailed numerical computations.

The frustrated magnetism has become an extremely developing field of research and one may anticipate further increases over the coming years. It should particularly be warned that the concept of frustration is often used, misused and even abused in the literature, and the widespread utilizations of this concept lead to many contradictions and misconceptions especially regarding residual entropy.

A lot of new exotic phenomena generated by frustrations have not yet been properly explained. There is not even a strict generally accepted definition of frustrations. And, as a result, neither a qualitative nor a quantitative measure of frustrations has been determined. Therefore, we believe that our paper will promote to understand some peculiar features in the spectacular field of frustrations in magnetism.

We also hope that our results may be and will be applicable to various lattices, Archimedean, Laves, and others that do not belong to these two classes, as well as to other models besides the Ising model.

\begin{acknowledgments}
The research was carried out within the state assignment of Ministry of Science and Higher Education of the Russian Federation (theme ``Quantum'' No.~122021000038-7). 
\end{acknowledgments}

\bibliographystyle{apsrev4-2}
\bibliography{frusts}

\end{document}